\documentclass[12pt]{article}

\usepackage{latexsym}

\newcommand{\sect}[1]{\setcounter{equation}{0}\section{#1}}

\begin{document}

\title{Trace anomaly of dilaton coupled scalars in two dimensions}
\author{{\sc Raphael Bousso}\thanks{\it R.Bousso@damtp.cam.ac.uk} \ 
  and {\sc Stephen W. Hawking}\thanks{\it S.W.Hawking@damtp.cam.ac.uk}
      \\[1 ex] {\it Department of Applied Mathematics and}
      \\ {\it Theoretical Physics}
      \\ {\it University of Cambridge}
      \\ {\it Silver Street, Cambridge CB3 9EW}
       }
\date{DAMTP/R-97/25\ \ \ \ hep-th/9705236 \\[1ex]
 submitted to Phys.\ Rev.\ Lett.}

\maketitle

\begin{abstract}
  
  Conformal scalar fields coupled to the dilaton appear naturally in
  two-dimensional models of black hole evaporation. We show that their
  trace anomaly is $(1/24\pi) [ R - 6 (\nabla \phi)^2 - 2 \Box \phi
  ]$. It follows that an RST-type coun\-ter\-term appears naturally in
  the one-loop effective action.

\end{abstract}

\pagebreak

\sect{Introduction}

In the study of black hole radiation, many useful results have been
obtained from two-dimensional (2D) models.  It is hoped that the
results will extend, at least partly, to the behaviour of realistic
black holes in four or more dimensions. To make this claim plausible,
the 2D actions were usually obtained by a dimensional reduction from a
higher-dimensional theory.  In the seminal papers of Callan, Giddings,
Harvey, and Strominger (CGHS)~\cite{CGHS}, and of Russo, Susskind, and
Thorlacius (RST)~\cite{RST1}, the action is
\[
S = - \frac{1}{16\pi G} \int d^2\!x\, \sqrt{-g}\, e^{-2\phi}
  \left( R + 4 (\nabla \phi )^2 + 4 \lambda^2 \right).
\]
This action is argued by CGHS to describe the radial modes of extremal
dilatonic black holes in four or more
dimensions~\cite{GidStr91,GarHor91}. Later, Trivedi and
Strominger~\cite{Tri93,StrTri93} studied a 2D model that was obtained
directly from 4D Einstein-Maxwell theory. A spherically symmetric
ansatz,
\[
ds^2 = g_{\mu\nu} dx^\mu dx^\nu + e^{-2\phi} d\Omega^2,
\]
allows the integration of the angular modes, and yields the action
\[
S = - \frac{1}{16\pi G} \int d^2\!x\, \sqrt{-g}\,  e^{-2\phi}
  \left( R + 2 (\nabla \phi )^2 + 2 e^{2\phi} - 
  2 Q^2 e^{4\phi} \right).
\]

For the description of black hole radiation, matter fields must be
included in the theory. For conformal matter, the trace of the
energy-momentum tensor vanishes classically; if treated as a quantum
field, however, the trace acquires a non-zero expectation value on a
curved background. The amount of radiation at infinity is directly
proportional to the trace anomaly~\cite{ChrFul77}. By including the
trace anomaly in the equations of motion, or, equivalently, by
including the one-loop effective action of the matter field, one can
study the back reaction of the evaporation on the geometry.

The simplest kind of matter that might be used is a minimally coupled
scalar field, $f$. To obtain a 2D model of genuine 4D pedigree, this
field must be included in the 4D theory. It must be reduced to 2D like
the rest of the action. Thus, the 4D classical action should be
amended by a term proportional to 
\[ 
\int d^4\!x\, \left(-g^{(4)} \right)^{1/2}
 \left( \nabla^{(4)} f \right)^2.
\]
By reducing this term to 2D, one finds that the
following matter contribution must be added to the 2D action:
\begin{equation}
S_{\rm m} = \frac{1}{2} \int d^2\!x\, \sqrt{-g}\,
  e^{-2\phi} (\nabla f)^2.
\label{eq-dilcoup}
\end{equation}
Thus, in the reduction process, the kinetic term acquires an
exponential coupling to the dilaton.

The next step is to calculate the trace anomaly or the one-loop
effective action, in order to include quantum effects. The field in
Eq.~(\ref{eq-dilcoup}) is conformally invariant; its trace anomaly,
however, was not known. Because of this problem, CGHS included the
field $f$ as a minimally coupled field in 2D:
\[
S_{\rm m} = \frac{1}{2} \int d^2\!x\, \sqrt{-g}\, (\nabla f)^2.
\]
For this field the trace anomaly is known, but its action could not
have arisen in the reduction from a realistic higher-dimensional
theory. This inconsistency pervades the entire literature on 2D
models. The problem was addressed most openly by Trivedi~\cite{Tri93},
who admitted that the 4D interpretation was lost when the minimal
scalars are introduced into the 2D model. This interpretational gap
seemed to become even wider when RST introduced a counterterm into the
effective action by hand, rendering the model solvable but even less
natural~\cite{RST1}.

In this paper we calculate the trace anomaly for the dilaton coupled
scalar field in two dimensions. This will make it possible to study
black hole radiation in 2D models that have a genuine 4D
interpretation. A particularly interesting result is that a
counterterm of the same form as that introduced by RST
appears naturally in the one-loop effective action for the
dilaton coupled scalar. We will use the zeta function approach,
together with general properties of the trace anomaly; a brief summary
of these methods is given in the following section. A more extensive
discussion is found, e.g., in Refs.~\cite{BirDav,Haw77}.

\sect{Methods}

From the eigenvalues $\lambda_n$ of the operator $A$, one defines a
generalised zeta function,
\[
\zeta(s) = {\rm tr} A^{-s} = \sum_n \lambda_n^{-s}.
\]
This sum converges for a sufficiently large real part of $s$. By
analytic extension, it defines a meromorphic function of $s$, which is
regular even in regions where the sum diverges.  The one-loop
effective action, $W$, is given by
\begin{equation}
W = \frac{1}{2} \left[ \zeta'(0) + \zeta(0) \log \mu^2 \right],
\label{eq-W-zeta'}
\end{equation}
where $\zeta' = d\zeta/ds$.
Under a rescaling of the operator,
\begin{equation}
A[k] = k^{-1} A,
\label{eq-A-scale}
\end{equation}
the one-loop effective action transforms as
\begin{equation}
W[k] = W + \frac{1}{2} \zeta(0) \log k.
\label{eq-W-k}
\end{equation}

We denote the trace anomaly by $T$. Let us summarise some of its
general properties in $D$-dimensional spacetimes~\cite{BirDav}. If $D$
is odd, the trace anomaly is zero.  If $D$ is even, it consists of
terms $T_i$ that are generally covariant and homogeneous of order $D$
in derivatives: $T = \sum q_i T_i$.  The dimensionless numbers $q_i$
are universal, i.e., independent of the background metric. This
property will be particularly useful, because it will allow us to
choose convenient backgrounds to determine the values of the $q_i$.

The integral of the trace anomaly over the manifold is given by:
\begin{equation}
\int d^D\!x \sqrt{g}\, T =
  2 \left. \frac{dW}{dk} \right|_{k=1},
\label{eq-T-W}
\end{equation}
where $k$ is defined as a scale factor of the metric,
$\hat{g}^{\mu\nu} = k^{-1} g^{\mu\nu}$. But under this scale
transformation, the eigenvalues of $A$ transform as in
Eq.~(\ref{eq-A-scale}). Therefore, Eq.~(\ref{eq-W-k}) can be used,
which yields the elegant result
\begin{equation}
\int d^D\!x \sqrt{g}\, T = \zeta(0).
\label{eq-T-zeta}
\end{equation}

Given an operator $A$, one could, in principle, calculate the one-loop
effective action directly from Eq.~(\ref{eq-W-zeta'}). In practice, it
is often simpler to calculate the trace anomaly from
Eq.~(\ref{eq-T-zeta}), because the zeta function is usually easier to
obtain than its derivative. By requiring that Eq.~(\ref{eq-T-W}) hold,
the effective action can be inferred up to terms that do not depend on
the scale factor. (We shall use $W^*$ to denote a quantity which
differs from the effective action only by such terms.) Also, if the
effective action is known for the operator $kA$, one can use
Eq.~(\ref{eq-W-k}) to obtain $W$ for the operator $A$.

\sect{Dilaton Coupled Scalar} \label{sec-dilcoup}

By Eq.~(\ref{eq-dilcoup}), scalar fields obtained through
dimensional reduction from four dimensions will have an $e^{-2\phi}$
dilaton coupling in the action. Variation with respect to $f$ yields
the equation of motion $Af=0$, with the field operator
\begin{equation}
A = e^{-2\phi} \left( - \Box + 2 \nabla^\mu \phi \nabla_\mu
\right).
\label{eq-A-dilcoup}
\end{equation}
The trace anomaly consists of covariant terms with two metric
derivatives. For the operator at hand, there are only three such
expressions: $R$, $(\nabla \phi)^2$, and $\Box \phi$.  In principle,
these terms could still be multiplied by arbitrary functions of
$\phi$. But consider shifting $\phi$ by a constant value $\Delta
\phi$. This corresponds merely to multiplying the kinetic term in the
action by a factor $e^{-2\Delta \phi}$; the trace anomaly will remain
the same. Therefore a functional dependence of any of its terms on
$\phi$ can be excluded. Consequently, we can write
\begin{equation}
T = q_1 R + q_2 (\nabla \phi)^2 + q_3 \Box \phi.
\label{eq-T-dilcoup}
\end{equation}
By writing the metric in conformal gauge,
\[
ds^2 = e^{2\rho(t,x)} \left( dt^2 + dx^2 \right),
\]
it is easy to check that this anomaly derives from the effective action
\begin{equation}
W^* = - \frac{1}{2} \int d^2\!x \sqrt{g}\, \left[
  \frac{q_1}{2} R \frac{1}{\Box} R +
  q_2 (\nabla \phi)^2 \frac{1}{\Box} R + q_3 \phi R \right].
\label{eq-W-dilcoup}
\end{equation}
This follows from Eq.~(\ref{eq-T-W}), since $R=-2\Box \rho$. (A more
straightforward result for the the last term would be $\Box \phi
\frac{1}{\Box} R$. It is related to the term we use by two
integrations by parts; the difference can at most be a boundary term.
It will become clear below why we choose the form $\phi R$.)  We must
only determine the universal numbers $q_1$, $q_2$, and $q_3$ to obtain
the trace anomaly completely.

\subsection{Coefficients of $R$ and of $\Box \phi$}

First consider the case when $\phi$ is identically zero. Then
Eq.~(\ref{eq-T-dilcoup}) simplifies to $T_{\phi \equiv 0} = q_1 R$.
But if $\phi \equiv 0$, the operator $A$ in Eq.~(\ref{eq-A-dilcoup})
becomes $A_{\phi \equiv 0} = - \Box$.  This is the operator for the
minimally coupled scalar, for which the trace anomaly is well
known~\cite{ChrFul77}: $T_{\rm min} = R/24\pi$.  Therefore, one finds
that
\[
q_1 = \frac{1}{24\pi}.
\]

Now consider the case where $\phi$ is constant, $\phi \equiv \phi_{\rm
  c}$. Then the one-loop
effective action, Eq.~(\ref{eq-W-dilcoup}), simplifies to
\begin{equation}
W^*_{\phi \equiv \phi_{\rm c}} = W^*_{\rm min}
  - \frac{1}{2} \int d^2\!x \sqrt{g}\,
  q_3 \phi_{\rm c} R.
\label{eq-W-phiconst}
\end{equation}
To make sure that the integral over the Ricci scalar does not vanish,
we can specify that a background with the topology of a two-sphere be
used.  For constant $\phi$, the operator $A$ becomes $A_{\phi \equiv
  \phi_{\rm c}} = - e^{-2\phi_{\rm c}} \Box$.  But this is just the
minimally coupled operator, rescaled by a constant factor $k^{-1} =
e^{-2\phi_{\rm c}}$.  Therefore, Eqs.~(\ref{eq-W-k})
and~(\ref{eq-T-zeta}) yield:
\begin{eqnarray}
W^*_{\phi \equiv \phi_{\rm c}} & = & W^*_{\rm min} + \phi_{\rm c}
  \zeta_{\rm min}(0) \nonumber \\
& = & W^*_{\rm min} + \phi_{\rm c} q_1 \int
  d^2\!x \sqrt{g}\, R.
\label{eq-Wmin-rescaled}
\end{eqnarray}
Comparison with Eq.~(\ref{eq-W-phiconst}) shows that
\[
q_3 = -2 q_1 = - \frac{1}{12\pi}.
\]

The same consideration also vindicates the choice of $\phi R$ for the
last term in the effective action, Eq.~(\ref{eq-W-dilcoup}): If $\Box
\phi \frac{1}{\Box} R$ was used, the last term in
Eq.~(\ref{eq-W-phiconst}) would be zero, since $\phi$ is constant. It
would then be impossible to match Eq.~(\ref{eq-W-phiconst}) to
Eq.~(\ref{eq-Wmin-rescaled}), in which the last term is non-zero on a
two-sphere background.\footnote{Nojiri and Odintsov~\cite{NojOdi97}
  suggest a more general form for the effective action, in which the
  last term is given by $ q_3 [ a \phi R + (1-a) \Box \phi
  \frac{1}{\Box} R ]$. This would give a different value of $q_3$.}

\subsection{Coefficient of $(\nabla \phi)^2$}

In conformal gauge the field operator will take the form
\[
A = e^{-2\phi-2\rho} \left[ -  \frac{\partial^2}{\partial t^2}
  - \frac{\partial^2}{\partial x^2}+ 2 \left(
  \frac{\partial\phi}{\partial t} \frac{\partial}{\partial t} +
  \frac{\partial\phi}{\partial x} \frac{\partial}{\partial x} \right)
  \right]. 
\]
Consider a Euclidean background manifold of toroidal topology, in
which $t$ and $x$ are periodically identified, with period $2\pi$. The
integral over the Ricci scalar is a topological invariant and vanishes
on a torus. Since $\Box \phi$ is a total divergence, its integral
vanishes as well. Thus,
\begin{equation}
\zeta(0) = \int d^2\!x \sqrt{g}\, T
   = q_2 \int d^2\!x \sqrt{g}\, (\nabla \phi)^2.
\label{eq-zeta-q2}
\end{equation}
Therefore we can determine $q_2$ by calculating $\zeta(0)$ from the
operator eigenvalues in a conveniently chosen toroidal background, and
dividing the result by $\int d^2\!x \sqrt{g}\, (\nabla \phi)^2$.

A useful choice of background is the field configuration $\phi = -\rho
= \epsilon \sin t$, where $ \epsilon \ll 1$. The operator
takes the form
\[
A = - \frac{\partial^2}{\partial t^2}
  - \frac{\partial^2}{\partial x^2}
  + 2 \epsilon \cos t \frac{\partial}{\partial t}.
\]
For $ \epsilon =0$, this operator is just the flat space Laplacian,
for which $\zeta(0)$ is known to vanish. The integral on the right
hand side of Eq.~(\ref{eq-zeta-q2}) yields $ 2 \pi^2 \epsilon^2$.
Thus we can proceed as follows: The eigenvalues of $A$ will be found
perturbatively in $ \epsilon $. This will allow us to expand
$\zeta(s)$ to second order in $ \epsilon $:
\[
\zeta(s) = \zeta^{(0)}(s)+ \epsilon \zeta^{(1)}(s)
  + \epsilon^2 \zeta^{(2)}(s).
\]
Since $\zeta^{(0)}(0)=0$, we have
\[
\epsilon \zeta^{(1)}(0) + \epsilon^2 \zeta^{(2)}(0) =
  2 \pi^2 q_2 \epsilon^2.
\]
Consistency requires that $\zeta^{(1)}(0)=0$; we will indeed find that
to be the case. Therefore,
\begin{equation}
q_2 = \frac{1}{2\pi^2} \zeta^{(2)}(0).
\label{eq-q2-zeta2}
\end{equation}

For $ \epsilon=0$, the eigenvalues of the operator $A$ are
$\Lambda^{(0)}_{kl} = k^2 + l^2$, with degeneracies
\[
d(k,l)= \left\{
  \begin{array}{ll}
  4 & \mbox{if $k \geq 1$, $l \geq 1$} \\
  2 & \mbox{if $k \geq 1$, $l=0$ or $k=0$, $l \geq 1$} \\
  1 & \mbox{if $k=l=0$.}
  \end{array} \right.
\]
Clearly, the zeta function, 
\[
\zeta(s) = \sum^{\infty}_{k,\,l=0}
d(k,l) \left( \Lambda^{(0)}_{kl} \right)^{-s},
\]
contains an ill-defined term: $k=l=0$. This problem can be dealt with
by introducing a mass term into the operator $A$: $A \rightarrow
A+M^2$. Then $\zeta(0)$ can be defined in the limit as $M \rightarrow
0$.

Now take $ \epsilon \neq 0 $, and consider the eigenvalue equation, $A
f = \Lambda f$. With $f(t,x) = T(t) X(x)$ the equation separates into
\[
- X'' = \sigma X,\ \ \ - \ddot{T} + 2 \epsilon \cos t\, \dot{T}
 = \lambda T.
\]
Standard perturbation theory yields that, to second order in $
\epsilon $, the eigenvalues of $A$ are
\[
\Lambda_{kl} = k^2 + l^2 + M^2 + \epsilon^2 \frac{2 l^2}{4 l^2 - 1},
\]
with the same degeneracies $d(k,l)$ as in the unperturbed case.
The zeta function is given by
\begin{eqnarray}
\zeta(s) & = & \sum^{\infty}_{k,\,l=0}
d(k,l) \left( \Lambda^{(0)}_{kl} \right)^{-s} \left( 1 + \epsilon^2
  \frac{\lambda^{(2)}_l}{\Lambda^{(0)}_{kl}} \right)^{-s}
  \nonumber \\
& = & \sum^{\infty}_{k,\,l=0}
d(k,l) \left( \Lambda^{(0)}_{kl} \right)^{-s} \left( 1 - \epsilon^2 s 
  \frac{\lambda^{(2)}_l}{\Lambda^{(0)}_{kl}} \right)
  \nonumber \\
& = & \zeta^{(0)}(s) - \epsilon^2 s
  \sum^{\infty}_{k=0,\,l=1} d(k,l) 
  \frac{\lambda^{(2)}_l}{\left( \Lambda^{(0)}_{kl} \right)^{1+s}},
\nonumber
\end{eqnarray}
where a Taylor expansion to second order in $ \epsilon $ was used. The
sum in the last line does not include $l=0$ because
$\lambda^{(2)}_0=0$. Since this excludes $k=l=0$, it is safe to drop
$M$ at this point. Thus we have
\begin{equation}
\zeta^{(2)}(0) = - \lim_{s \rightarrow 0} s U(s),
\label{eq-zeta2-U}
\end{equation}
where we view the double sum as a meromorphic function of $s$,
\[
U(s) = 
  \sum^{\infty}_{k=0,\,l=1} d(k,l) 
  \frac{2 l^2}{(k^2+l^2)^{1+s} (4 l^2 - 1)}.
\]
If $U(0)$ were finite, $\zeta^{(2)}(0)$ would vanish; but it is easy
to check that this is not the case. If $U$ had poles of order 2 or
greater, $\zeta^{(2)}(0)$ would diverge. Only if $U$ has a simple pole
at $s=0$ will we obtain a non-zero, finite result for
$\zeta^{(2)}(0)$, and thus for $q_2$. We show below that this is
indeed the case.

To understand fully the behaviour of $U(s)$, we would have to evaluate
it in terms of known meromorphic functions. Fortunately we need to
find only the principal part of the Laurent series of $U$ around
$s=0$, ${\rm Pr} [U(s);0]$, because the regular part will be annulled
by the factor of $s$ in Eq.~(\ref{eq-zeta2-U}). But ${\rm Pr} [U(s);0]
= {\rm Pr} [U(s)+V(s);0]$ for any function V(s) which is regular at
$s=0$. Thus, by adding suitable finite terms to the double sum, we can
bring it into a form which can be evaluated.

First, we note that the contribution from $k=0$ is finite at $s=0$:
\[
2 \sum^{\infty}_{l=1} \frac{2}{4 l^2 - 1} = 2.
\]
After its subtraction, all summations start at 1:
\[
{\rm Pr} [U(s);0] = 2\, {\rm Pr} \left[
  \sum^{\infty}_{k,\,l=1}
  \frac{4 l^2}{(k^2+l^2)^{1+s} (4 l^2 - 1)} ; 0 \right],
\]
where we have used $d(k,l)=4$.  Next, we subtract 1 in the numerator;
this is possible since $ \sum (k^2+l^2)^{-1-s} (4 l^2 - 1)^{-1} $ is
finite at $s=0$ (the upper bound $-1+\frac{\pi^2}{12} + (\log 2 -
\frac{1}{2}) \pi \coth \pi \approx 0.43$ is easily found). The
cancellation of terms yields
\[
{\rm Pr} [U(s);0] = 2\, {\rm Pr} \left[
  \sum^{\infty}_{k,\,l=1}
  \frac{1}{(k^2+l^2)^{1+s}} ; 0 \right].
\]
But
\[
\sum^{\infty}_{k,\,l=1} \frac{1}{(k^2+l^2)^{1+s}} = 
\frac{1}{4} Z_2(2+2s) - \zeta_{\rm R}(2+2s),
\]
where
\[
Z_2(p) = \sum^{\infty}_{k,\,l=-\infty}
  \!\!\!\!\!^\prime\ (k^2+l^2)^{-p/2}
\]
is a generalised zeta function of Epstein type; the prime denotes the
omission of the $k=l=0$ term in the sum. Epstein showed in
Ref.~\cite{Eps03} that $Z_2(p)$ is analytic except for a simple pole
at $p=2$, with residue $2\pi$. Since the Riemann zeta function
$\zeta_{\rm R}(2+2s)$ is finite for $s=0$, we find
\[
2 {\rm Pr} [U(s);0] = \frac{1}{2}\, {\rm Pr}
      \left[ Z_2(2+2s);0 \right]
  = \frac{1}{2} \left( \frac{2\pi}{2s} \right)
  = \frac{\pi}{2s}.
\]
Therefore, by Eq.~(\ref{eq-zeta2-U}), we find that $\zeta^{(2)}(0) =
-\pi/2$, and, by Eq.~(\ref{eq-q2-zeta2}), we obtain the result
\[
q_2 = - \frac{1}{4\pi}.
\]

\sect{Summary}

We have shown that a 2D conformal scalar field with
exponential dilaton coupling has the trace anomaly
\[
T = \frac{1}{24\pi} \left[ R - 6 (\nabla \phi)^2 - 2 \Box \phi
\right].
\]
The scale factor dependent part of the one-loop effective action is
\[
W^* = - \frac{1}{48\pi} \int d^2\!x \sqrt{g}\, \left[ \frac{1}{2}
  R \frac{1}{\Box} R - 6 (\nabla \phi)^2 \frac{1}{\Box} R
  - 2 \phi R \right].
\]
This is the proper one-loop term that should be used in 2D
investigations of black hole evaporation. It is interesting to note
that the last term was inserted by hand in the RST model, albeit with
a different coefficient. By Eq.~(\ref{eq-W-zeta'}), the effective
action will also contain a term $(1/2) \zeta(0) \log \mu^2$. The $R$
and $\Box \phi$ terms in the trace anomaly give only a topological
contribution to $\zeta(0)$, which does not affect the equations of
motion. The term 
\[
- (1/8\pi)\log \mu^2 \int d^2\!x \sqrt{g}\, (\nabla
\phi)^2,
\]
however, must be taken into account.

The minimally coupled scalars, which are usually included in the 2D
theory to carry the black hole radiation, have no higher-dimensional
interpretation.  With our result, it will be possible to
investigate, for the first time, 2D models derived entirely from
higher-dimensional theories.

\section*{Acknowledgements}

We thank G. Gibbons and M. Perry for discussions. We are grateful to
S. Nojiri, S. Odintsov, D. Vassilevich, and E. Elizalde for their
comments on an earlier version of this letter.

\end{document}